\documentclass{article}

\usepackage[T1,T2A]{fontenc}
\usepackage[english]{babel}
\usepackage{graphicx,parskip,appendix,float}
\usepackage[ruled] {algorithm2e}
\usepackage{multirow}
\usepackage{authblk}
\usepackage{url,amsmath,amssymb,fancybox,listings,pdfpages,caption,multicol,datetime,rotating, booktabs}
\usepackage[pagebackref=false,pdffitwindow=true]{hyperref}
\usepackage{csquotes}
\usepackage[backend=biber, sorting=none]{biblatex}
\begin{document}

\title{Noise-tolerant tomography of multimode
linear optical interferometers with single photons}
\author[1,*]{Yu.A. Biriukov}
\author[1]{R.D. Morozov}
\author[1,2]{I.V. Dyakonov}
\author[3]{M.V. Rakhlin}
\author[3]{A.I. Galimov}
\author[3]{G.V. Klimko}
\author[3]{ S.V. Sorokin }
\author[3]{ I.V. Sedova}
\author[3]{ M.M. Kulagina}
\author[3]{ Yu.M. Zadiranov}
\author[3]{ A.A. Toropov}
\author[1]{A.A. Korneev}
\author[1]{S.P. Kulik}
\author[1,2]{S.S. Straupe}

\affil[1]{Quantum Technology Centre and Faculty of Physics, M.V. Lomonosov Moscow State University, 1 Leninskie Gory, Moscow 119991, Russia}
\affil[2]{Russian Quantum Center, 30 Bolshoy bul'var building 1, Moscow 121205, Russia}
\affil[3]{Ioffe Institute, 26 Polytekhnicheskaya street, St. Petersburg 194021, Russia}
\affil[*]{biriukov.ia18@physics.msu.ru} 
\date{}
\maketitle

\begin{abstract}
Linear optical networks are fundamental to the advancement of quantum technologies, including quantum computing, communication, and sensing. The accurate characterization of these networks, described by unitary matrices, is crucial to their effective utilization and scalability.  In this work, we present the method for reconstructing the transfer matrix of a linear optical interferometer based on the analysis of cross-correlation functions of photon counts between pairs of output modes. Our approach accounts for losses and photon indistinguishability, making it robust to experimental imperfections. By minimizing the requirements for the input states, the method simplifies the experimental implementation. We demonstrate the effectiveness of our technique through theoretical modeling and experimental validation in a 4-mode programmable integrated optical interferometer. The results show high fidelity in matrix reconstruction and successful application in boson sampling experiments. In addition, we provide a comprehensive formalism for correlation functions and discuss the robustness of the method to measurement errors. This work offers a practical and efficient solution for characterizing linear-optical networks, paving the way for scaling up photonic quantum technologies.
\end{abstract}

\section{Introduction}

Linear optical interferometers are foundational to quantum photonic technologies, enabling precise manipulation of quantum states of light for applications such as quantum computing, communication, and sensing~\cite{wang2020integrated, roadmap2022}. These systems are mathematically described by $N\times N$ unitary matrix $U$, where each element $U_{ij}$ encodes the amplitude and phase transformation from input mode $i$ to output mode $j$. Accurate reconstruction of $U$ is critical for verifying quantum operations, yet real-world imperfections -- photon loss, phase errors, mode mismatch, and detector inefficiencies -- complicate this task. These challenges grow with system size, demanding scalable and robust characterization methods.

Traditional approaches like quantum process tomography (QPT)~\cite{mohseni2008quantum, lobino2008complete, keshari2010quantum} provide full process descriptions, including transmission matrix $U$, but require exponentially scaling measurement set, which makes them impractical for large N. So, there is a need for more specific approaches that require fewer resources. Recent advances aim to reduce complexity: some approaches use classical laser light \cite{rahimi2013direct, volkoff2024learning, Smith:24} and achieve $O(N)$ numbers of measurements but depend on precise input-state control, while others rely on measurement single-photon interference and its analysis~\cite{laing2012super, hoch2023characterization, spagnolo2017learning, dhand2016accurate, tillmann2016unitary} that are insensitive to phase fluctuations and nonuniform losses but require indistinguishability tuning of incident photon pairs, which doubles total number of measurements compared to approaches with classical light. Neither fully addresses scalability and both require significant modifications in the experimental setup to measure $U$. 

In contrast, our approach does not require any significant change in the experimental setup: we use the analysis of time-resolved cross-correlation functions of photon counts at the outputs, which removes the need to modify indistinguishability and reduces the required number of measurements to \(O(N)\) while maintaining high reconstruction fidelity. Our method accounts for non-uniform losses in interferometer modes and photon indistinguishability, making it more robust to experimental and source imperfections.  

We benchmark it against two previous approaches that are similar to ours: 1) Laing and O’Brien’s super stable tomography~\cite{laing2012super} and Tillman’s optimization-based reconstruction algorithm~\cite{tillmann2016unitary}. Both approaches are based on photon interference and coincidence measurements with distinguishable and indistinguishable photons. The first one uses an analytical approach and is very sensitive to noise in experimental data, while the second one is based on optimization and shows robustness to noise and high performance in terms of fidelity between reconstructed and original matrices. As we demonstrate further, our method achieves fidelity and noise tolerance (up to 20\% error resilience) comparable to Tillman’s approach but with half the measurements needed. This efficiency stems from leveraging cross-correlation histograms, which encode both module and phase information through photon timing, bypassing the need for distinguishable photon input.

In the following sections, we will present the simplest theoretical description of our approach, results of analysis of algorithm performance, and experimental implementation of this approach for 4-mode linear-optical interferometers. The detailed theoretical description of the reconstruction method can be found in the Appendix ~\ref{Cor_func_math}.
\section{Algorithm description}
\subsection{Theoretical model}
\begin{figure}[ht]
\centering
\includegraphics*[width=1.\linewidth]{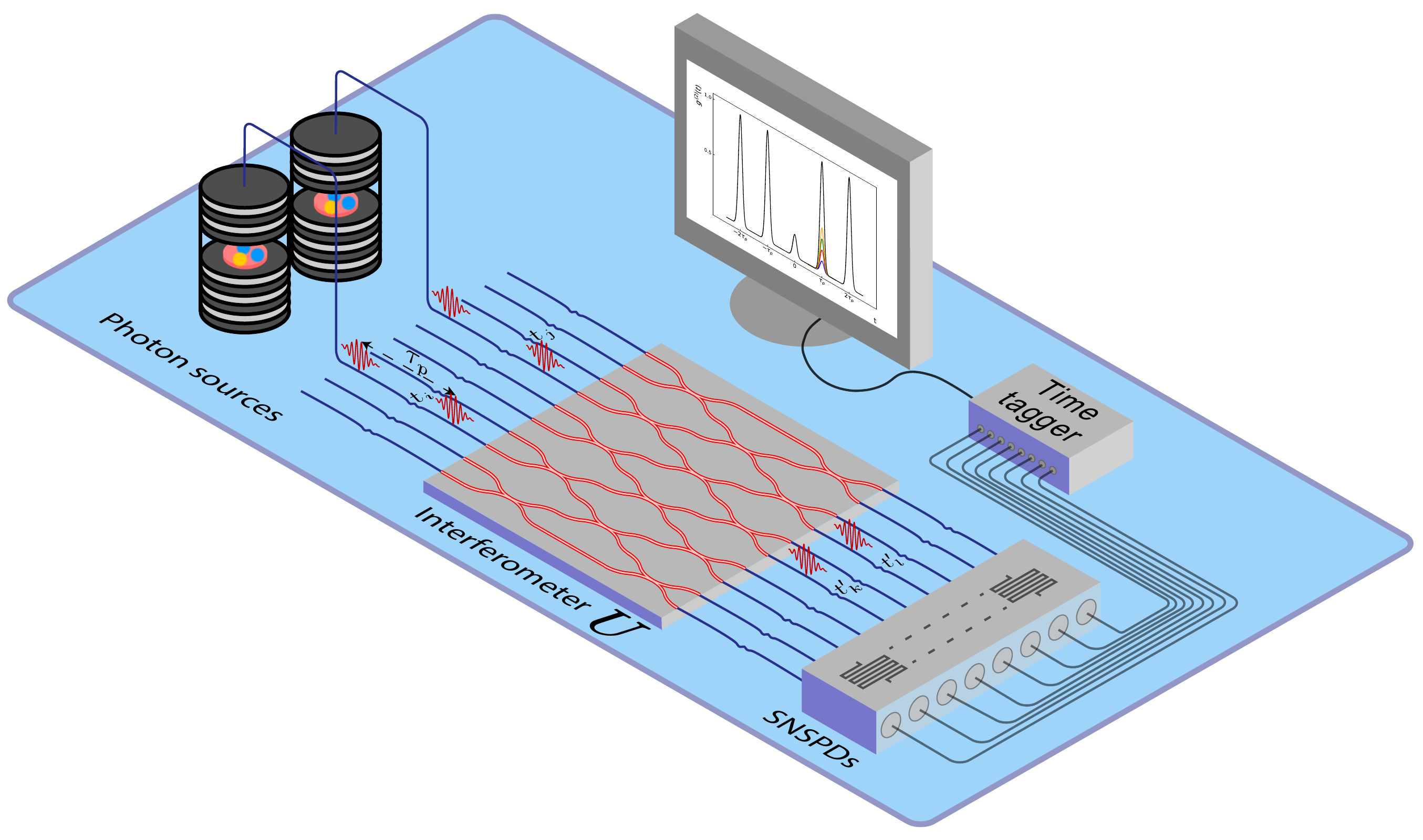}
\caption{Schematic diagram of the experimental setup. \(t_i, t'_j\) are the transmission coefficients in the \(i\)-th input and \(j\)-th output modes of the interferometer respectively, \(U\) is the transfer matrix of the interferometer, \(\tau_p\) is the period of the photon source pump pulses. On the screen there is schematic diagram of the measured cross-correlation function. Different colors show the contributions of different pulse pairs to the side peak on the histogram.}
\label{Setup}
\end{figure}

Our proposed method for tomography of the transfer matrix of a multimode linear-optical interferometer is based on measuring the cross-correlation function of photon counts between pairs of output modes when two single photons are periodically injected into different pairs of input modes of the interferometer. We focus on the setups with pulsed single-photon source, because these provide intuitive synchronization capabilities required for scalable quantum photonic experiments. The cross-correlation histograms obtained in the experiment represent periodic peaks of varying areas (see the diagram on the Fig. \ref{Setup}). 
The period of the peaks coincides with the period of the source pump pulses, and their areas are determined second-order correlation function of photon counts in the output modes $k$ and $l$:
\begin{equation}
G^{(2)}(t_1,t_2) = \langle a_k^\dagger(t_1)a_l^\dagger(t_2)a_k(t_1)a_l(t_2)\rangle,
\end{equation}
where $a_k(t)$ and $a_l(t)$ are the annihilation operators for photons in modes $k$ and $l$ at time $t$. Using the Heisenberg picture, we express these operators in terms of input modes $i$ and $j$ through the non-unitary matrix $M=T'UT$, which accounts for losses via diagonal transmission matrices $T_{ij} = t_i\delta_{ij},T' =  t'_{i}\delta_{ij}$, where coefficients $t_i$ include all photon loss channels before the interferometer (coupling into the fiber, coupling between the fiber modes and the interferometer modes, etc.) and the $t'_{j}$ include loss channels after the interferometer (coupling between the interferometer modes and the output fiber modes, quantum efficiencies of the detectors, etc.). The submatrix $M'$ that governs the transformations between the input pair $(i,j)$ and the output pair $(k,l)$ is:

\begin{equation}
M'=\begin{pmatrix}
M_{ki} & M_{kj}\\
M_{li} & M_{lj}
\end{pmatrix}
\end{equation}

Using cross-correlation formalism it can be shown (see Appendix \ref{Cor_func_math} for derivation) that for a pair of single photons $|1\rangle_{i}$, $|1\rangle_{j}$ with indistinguishability $I$ entering modes $i$ and $j$ the integrated areas of the central $\langle A_0 \rangle$ ($t_1 = t_2$) and side $\langle A_k \rangle$ ($t_1 = t_2 + k\tau_p$) peaks on the cross-correlation histogram are proportional to:
\begin{align}
\begin{split}
&\langle A_0 \rangle=\frac{1+I}{2}|Per(M')|^2 + \frac{1-I}{2}|Det(M')|^2, \\
&\langle A_k \rangle =|M'_{11}|^2|M'_{21}|^2 + |M'_{12}|^2|M'_{22}|^2+|M'_{11}|^2|M'_{22}|^2 + |M'_{12}|^2|M'_{21}|^2
\end{split}
\label{formula:area}
\end{align}
where $Per(M')$ and $Det(M')$ are the permanent and determinant of $M'$. The expression for $\langle A_0 \rangle$ is clear, while the $\langle A_k \rangle$ expression can be interpreted in the following way. Since before the interferometer single-photon wave packets do not interfere, the area of a particular peak can be considered as the sum of independent contributions from different pulse combinations. Different colors in Fig. \ref{Setup}, illustrate the contribution of various combinations to the peak areas. 

Next, we introduce a quantity that we will call "visibility" \(V_{ij}^{kl}\), as the ratio of the area of the central peak \(A_0\) to the average height of the side peaks \(\langle A_k \rangle\). 
\begin{equation}
V^{kl}_{ij} =\frac{A_0}{\langle A_i \rangle} = \frac{\frac{1+I}{2}|Per(U')|^2 + \frac{1-I}{2}|Det(U')|^2}{|U'_{11}|^2|U'_{22}|^2 + |U'_{12}|^2|U'_{21}|^2 + x_{ij}|U'_{11}|^2|U'_{21}|^2 + \frac{1}{x_{ij}}|U'_{12}|^2|U'_{22}|^2},
\label{Visib}
\end{equation}
where \(U'\) is the submatrix of \(U\) defined by the same elements as \(M'\), and \(x_{ij} = |t_i/t_j|^2\). Its advantage is that it does not depend on the histogram accumulation time (both areas of central and side bands depend linearly on it) or the output losses \(t'_k, t'_l\), while it depends only on the ratio of input losses. Thus, the experimentally measured visibilities \(\{V_{exp}\}\) can be compared with the theoretical \(V_{theor}\) calculated using formula \ref{Visib}.
\subsection{Optimization and complexity}
\begin{figure}[ht]
\centering
\includegraphics*[width=.5\linewidth]{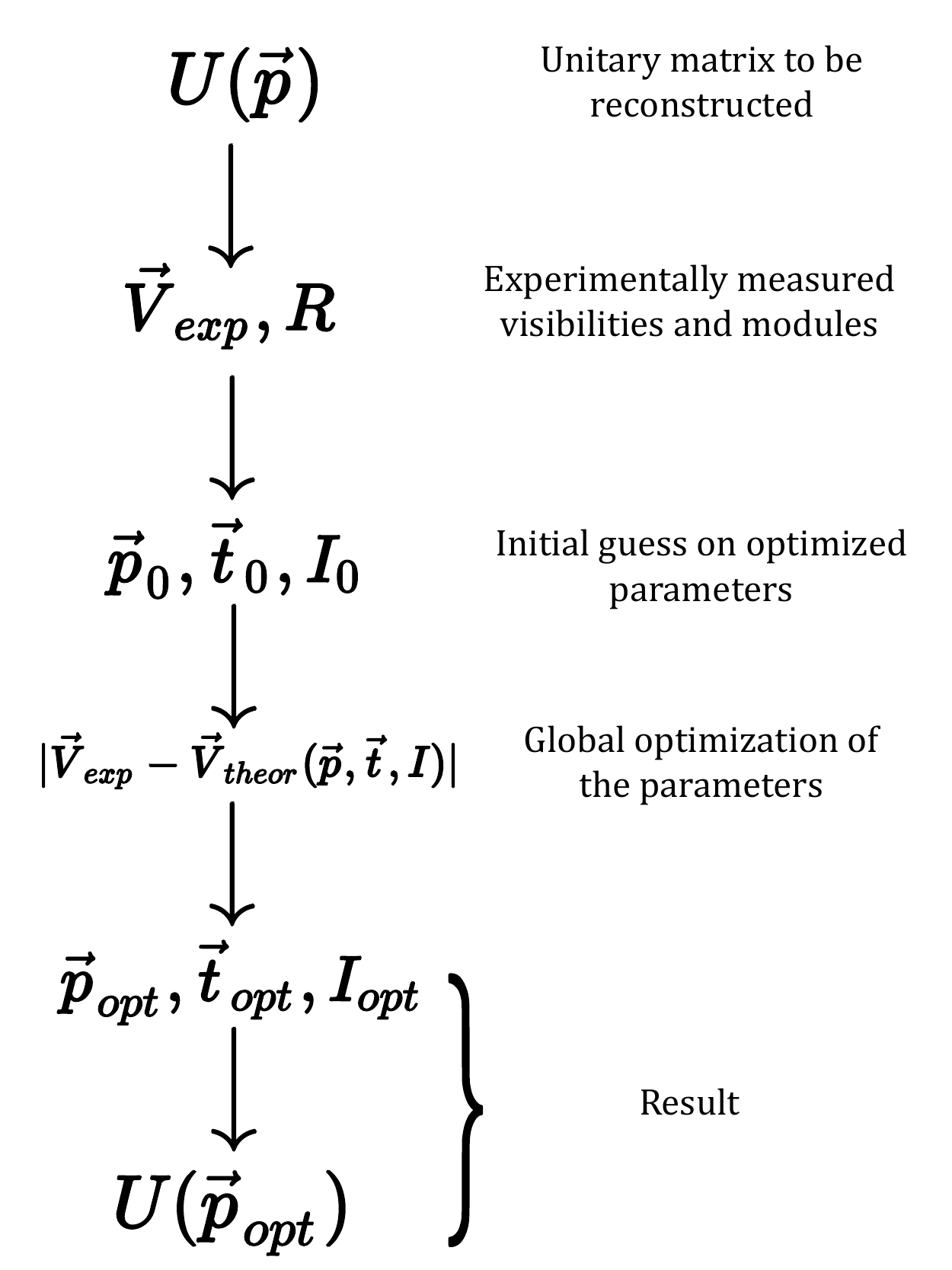}
\caption{Flowchart for our reconstruction algorithm. Primary data $\vec{V}_{exp}, R$ from a interferometer described by the unitary matrix $U(\vec{p})$ is measured. In general this
data will be error afflicted. The layout and initial parameters $\vec{p}, \vec{t}, I$, are either known apriori or are obtained by a reconstruction described in main text. These initial parameters are now subjected to a global optimization using an, in the best case over-complete, set of primary data. Here the output yields both, the reconstructed unitary $U(\vec{p}_{opt}) $and the parameters of the individual building blocks $\vec{p}_{opt}, \vec{t}_{opt}, I_{opt}$.}
\label{Flowchart}
\end{figure}

The further reconstruction procedure is as follows (see Fig.\ref{Flowchart}). 
A vector $\vec{p}$ parametrizes the desired transfer matrix $U$ according, for example, to Reck \cite{PhysRevLett.73.58} decomposition. The input transmission coefficients $\vec{t}$ and the indistinguishability of the photons $I$ also serve as parameters. For each such set of parameters, it is possible to calculate the theoretical visibilities \(V_{theor}(\vec{p}, \vec{t}, I)\) using \ref{Visib}, which can then be compared with \(V_{exp}\). From the set of $V_{exp}$ and some additional single-photon measurements the initial guess for the model parameters can be calculated (see \ref{section: Simulation} for details). A global optimization algorithm is launched, selecting the set of parameters that minimizes the difference between the theoretically calculated and experimentally obtained visibilities:
\begin{equation}
    \vec{p}_{opt}, \vec{t}_{opt}, I_{opt} = \underset{\vec{p}, \vec{t}, I}{argmin} \sum_{ijkl}{([V_{theor}(\vec{p}, \vec{t}, I)]_{ij}^{kl} - [V_{exp}]_{ij}^{kl})^2}
\end{equation}
The optimal parameter set $\vec{p}_{opt}$ defines the reconstructed interferometer matrix. It is clear that the larger is the visibility dataset the more accurate reconstruction should be expected. For small interferometers it is possible to measure visibility value for each combination of input and output channels.  It requires measurement of cross-correlation between all output channels for $M(M-1)/2$ pairs of input channels, which may be difficult for larger interferometers with tens of modes or more. In the Section \ref{section: Simulation} we show that based on work \cite{laing2012super} only $2\times M$ pairs of input channels are required with additional $M$ single-photon measurements to build a proper initial guess which is fed to the optimization algorithm together with same measured dataset. The optimization routines refines the reconstructed interferometer matrix. So in terms of the number of required measurements, complexity of the reconstruction procedure scales linearly with the number of modes of interferometer, since we measure all cross-correlations for fixed input simultaneously.   
\section{Numerical modeling}
\label{section: Simulation}

We numerically compared the efficiency of our proposed algorithm for reconstructing the unitary matrix of an interferometer with the SST method \cite{laing2012super} and an optimization algorithm based on \cite{laing2012super} proposed in \cite{tillmann2016unitary} and modified to take into account the indistinguishability of photons using formula \ref{formula:area} for the area of the central peak. 

The numerical reconstruction procedure is as follows. First, a Haar random unitary \(U\) and diagonal matrices of input and output mode transmissions of the interferometer \(T = t_i\delta_{ij}\), \(T' = t'_i\delta_{ij}\) are generated, where \(t_i^2 {t'_j}^2\) take random values from a uniform distribution on the interval \([0.5, 1]\) (although it will work with every nonzero indistinguishability). For a set of combinations of the input and output channel numbers of the interferometer, interference visibilities are generated using the theoretical formula \ref{Visib}. In simulations, the indistinguishability value was chosen to be $0.9$, which is well correlated with the experimentally measured value \ref{HOM_dip}. Furthermore, in the experimental conditions, the data set will be subject to noise. This perturbation of the initial data is simulated by adding noise from a normal distribution \((\tilde{V}, \tilde{R}) = (1 + N(0, \sigma^2))(V, R)\).

For our reconstruction method and for SST \cite{laing2012super}, the transfer matrix of the interferometer \(R = |M|^2\) is also required. It can be obtained by measuring single photon counts at the outputs of the interferometer if a photon stream is coupled only to a single input mode \cite{laing2012super}. 

Before starting the global optimization procedure, an initial guess \(U_0\) of the unitary matrix of the interferometer is required. For this, we use the SST method \cite{laing2012super}, as it does not require additional measurements. However, the SST method requires visibilities $V'$ measured in the scheme with distinguishable and indistinguishable photons \cite{laing2012super}, so in our approach, it is necessary to bring \(\tilde{V}\) introduced here to \(\tilde{V}'\) from \cite{laing2012super}. This is done using the formula (assuming that the value of indistinguishability is $1$):

\begin{equation}
\tilde{V'}^{kl}_{ij} = 1 + \left(1 + \alpha^{kl}_{ij}\right)\left(\tilde{V}^{kl}_{ij} - 1\right)
\end{equation}
\begin{equation}
\alpha^{kl}_{ij} = \frac{\tilde{x}_{kl}|\tilde{U_{ik}}\tilde{U_{jk}}|^2 + \tilde{x}_{kl}^{-1}|\tilde{U_{il}}\tilde{U_{jl}}|^2}{\tilde{|U_{ik}}\tilde{U_{jl}}|^2 + |\tilde{U_{il}}\tilde{U_{jk}}|^2}
\end{equation}

where \(\tilde{x}_{kl} = \tilde{t}_k^2 \tilde{t}_l^2\) and \(A_{ij} = |\tilde{U}_{ij}|^2\) are approximations of the transmission ratios in the output channels and the matrix of squared magnitudes of the unitary matrix of the interferometer, respectively. These quantities can be calculated using the Sinkhorn-Knopp algorithm \cite{sinkhorn1964relationship}, which selects diagonal matrices \(D_{kl} = \tilde{t}_k^2 \delta_{kl}\), \(D'_{kl} = \tilde{t'}_k^2 \delta_{kl}\) that bring \(\tilde{R}\) to a doubly stochastic form: \(\tilde{R} = D'AD\). The desired unitary matrix is parameterized via the Reck decomposition \(\tilde{U}(\tilde{p})\) \cite{PhysRevLett.73.58}. Thus, the task of the global optimizer is to select a set of parameters \((\tilde{p}, \tilde{x}, \tilde{I})\) that minimizes the difference between the experimentally measured interference visibilities and the theoretical prediction. The whole procedure is summarized on Fig. \ref{Flowchart}.

The simulation results are shown in Fig. \ref{Simulation_modes}. The number of visibilities to reconstruct the matrix was equal for both methods from \cite{tillmann2016unitary} and this work. It can be seen that the best result is provided by the method from \cite{tillmann2016unitary}, however, it requires twice as many measurements as the method proposed in this article, because it requires visibilities from \cite{laing2012super}. Moreover, the model used in \cite{tillmann2016unitary} does not include indistinguishability of photons. Thus, our method is a good compromise between the quality of reconstruction and the number of measurements required to reconstruct the matrix in a real-world scenario.

\begin{figure}[ht]
\centering
\includegraphics*[width=.8\linewidth]{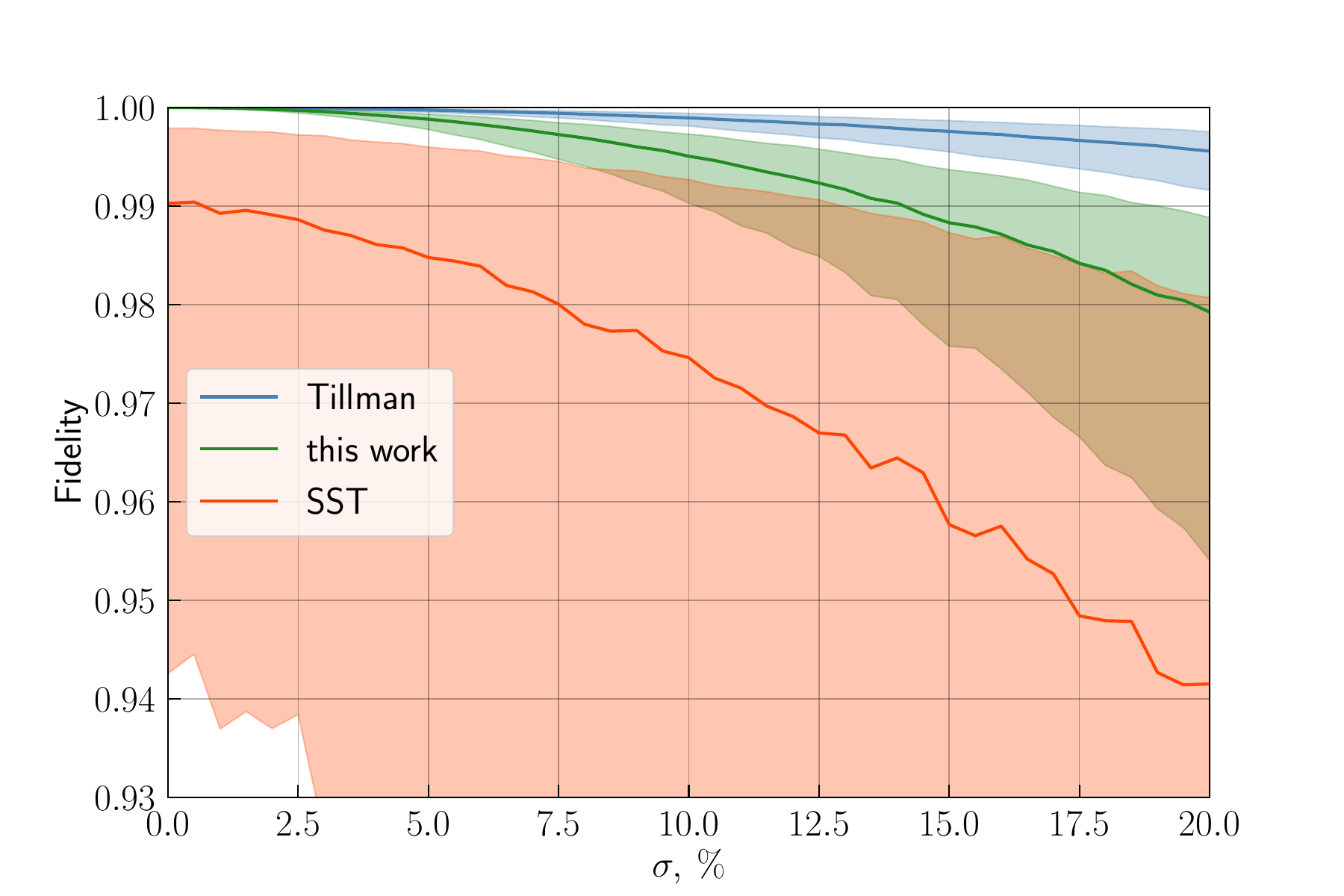}
\caption{Simulation results of the dependence of reconstruction quality on noise level $\sigma$, which is occurs in the regime of big dark count rate compared to signal counts. Errors are 90\% confidence intervals. Three methods are presented: SST—the method proposed by \cite{laing2012super}, Tillman—-the optimization method from \cite{tillmann2016unitary}, this work-—the method proposed in this article.}
\label{Simulation_noise}
\end{figure}

We also simulated how the quality of reconstruction changes with the number of modes and, surprisingly, find that it gets better (see Fig. \ref{Simulation_modes}) while the results of SST \cite{laing2012super} are getting worse. Noise level was fixed to be 10\% of the absolute value of visibilities. It can be understood from the simple comparison of number of parameters in model and number of measured values. In Reck decomposition \cite{PhysRevLett.73.58} that we use there are $O(M^2)$ phases of inner $2\times2$ Mach-Zender interferometers. Meanwhile, we measure $M(M-1)/2$ visibilities for $2*M$ pairs of input channels, that comprises $O(M^3)$ visibilities. The number of measured values grows faster than the number of undefined parameter in our model, so it makes the quality of reconstruction higher.

\begin{figure}[ht]
\centering
\includegraphics*[width=.8\linewidth]{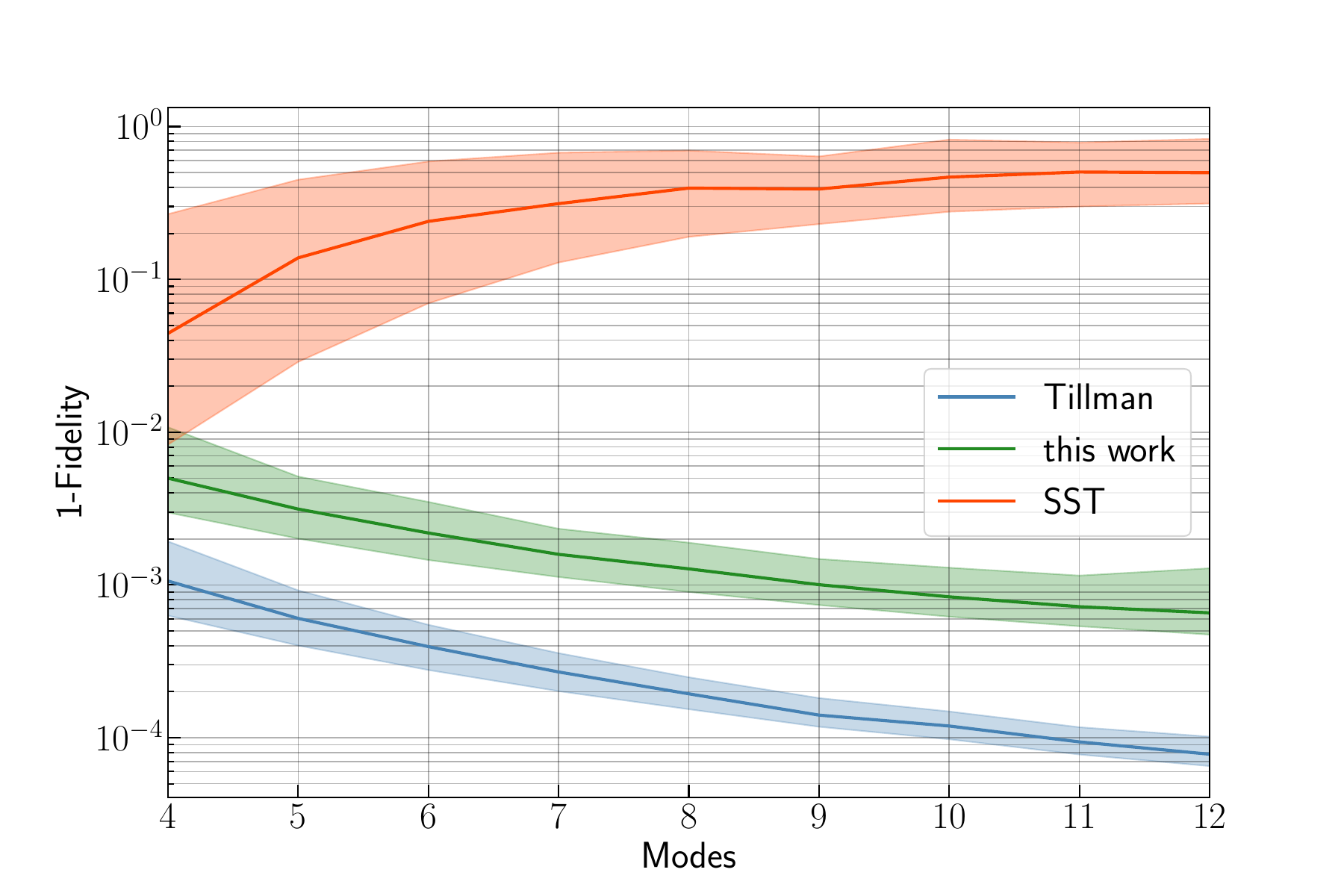}
\caption{Simulation results of the dependence of reconstruction quality on number of modes of the interferometer. Errors are 90\% confidence intervals. Three methods are presented: SST—the method proposed by \cite{laing2012super}, Tillman-—the optimization method from \cite{tillmann2016unitary}, this work-—the method proposed in this article.}
\label{Simulation_modes}
\end{figure}
\section{Experimental results}
\subsection{Matrix reconstruction}
\begin{figure}[ht]
\centering
\includegraphics*[width=1.\linewidth]{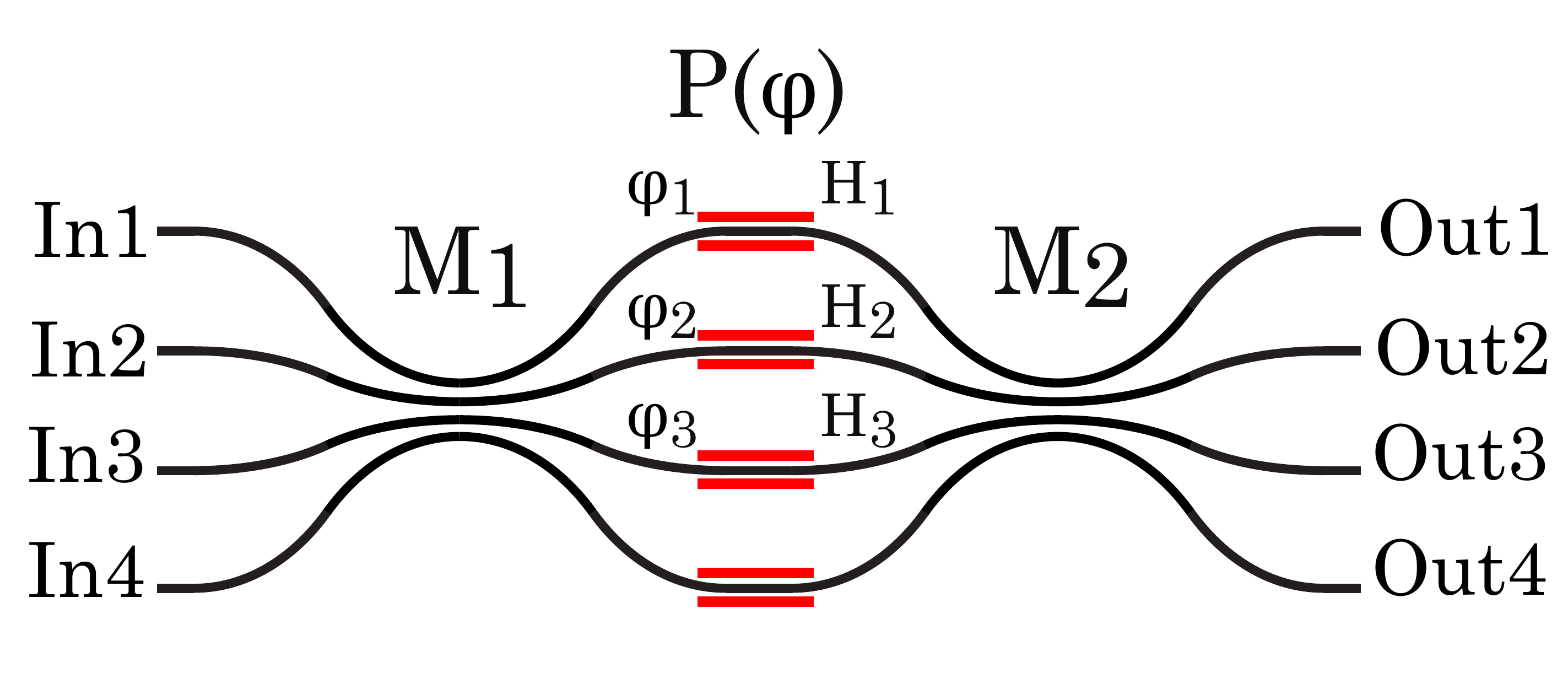}
\caption{Schematic optical chip structure. There are two mixing layers $M_1, M_2$ and programmable phase shifter layer $P(\phi)$ between them. Phases $\phi_1,\phi_2,\phi_3$ are calibrated and programmed using thermooptical effect.}
\label{Chip_struct}
\end{figure}

The reconstruction algorithm was implemented in the experiment with a 4-mode programmable integrated-optical interferometer. The chip was fabricated using femtosecond laser writing technique in fused silica \cite{skryabin2024femtosecond} and represents two mixing layers and one phase layer between them of robust architecture \cite{saygin2020robust} (see Fig \ref{Chip_struct}). 
The programmability of the interferometer is enabled by the use of thermo-optic phase shifters. A semiconductor QD with emission probability \(p_{emit} \approx 0.15\%\) was used as the single-photon source at 919 nm. The single photons from QD were demultiplexed into two channels with appropriate temporal delays used for photon synchronization. The complete chip model had already been reconstructed using an algorithm similar to the one proposed in the article - the obtained matrix was used as a reference \cite{Ilya}.

\begin{figure}[ht]
\centering
\includegraphics*[width=1.\linewidth]{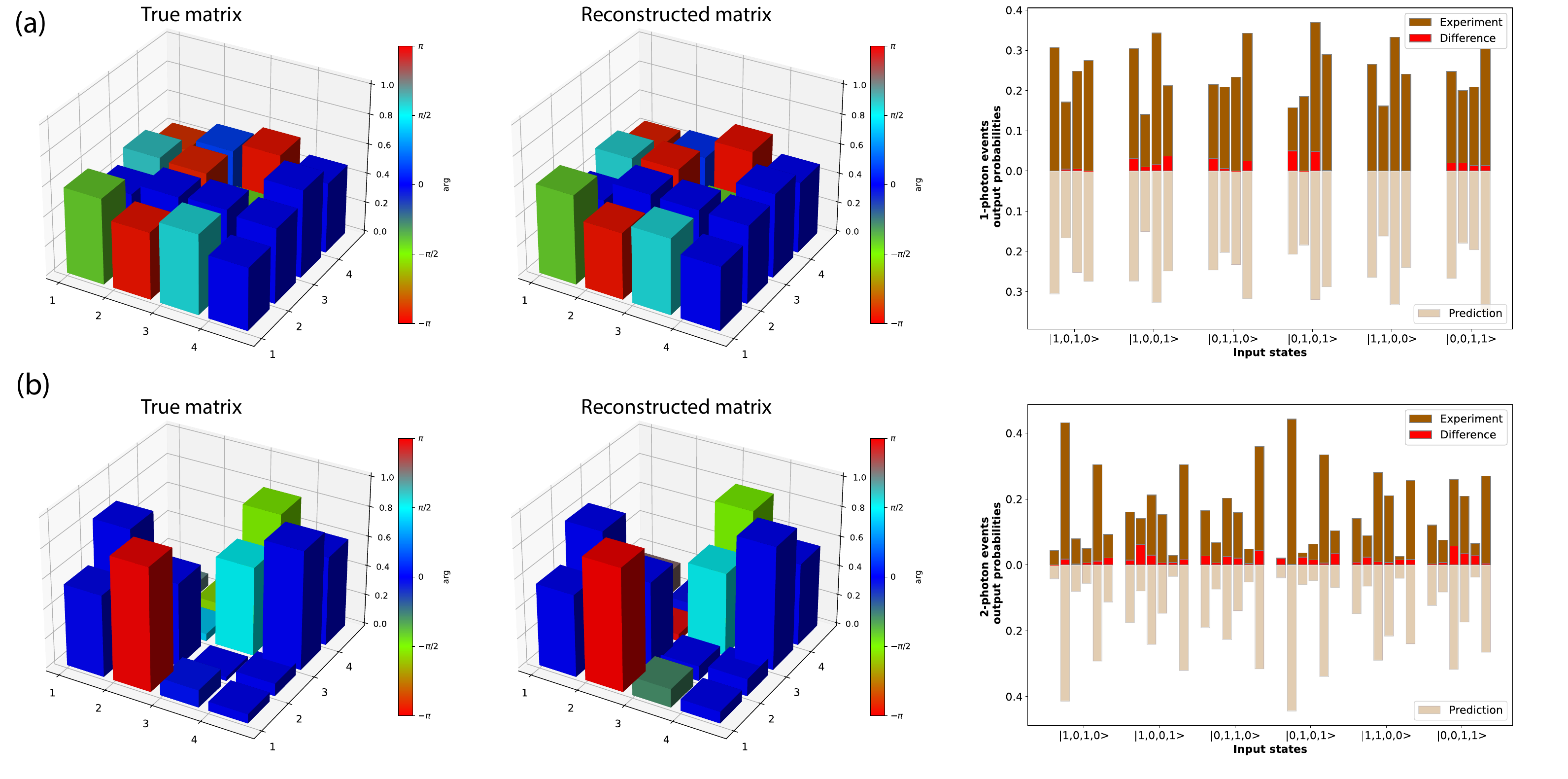}
\caption{Experimental implementation of proposed algorithm for transfer matrix tomography. (a-b) Results of reconstructing a transfer matrices of 4-mode integrated interferometer: with approximately equal magnitudes (a) and with half of the elements close to zero (b). The height of the bars shows the magnitudes of the elements, and the color shows the phases. (c-d) Results of boson sampling. Comparison of experimental and simulation results for single-photon (c) and two-photon (d) events at the chip outputs.}
\label{Exp_results}
\end{figure}

The reconstruction algorithm tests were carried out for two different configurations of currents were applied to the photonic circuit to obtain two different transfer matrices. The first one results in a matrix with random magnitudes and phases (see Fig. \ref{Exp_results} (a)), and the second one in the matrix with approximately half of the elements close to zero (see Fig. \ref{Exp_results} (b)), to test the algorithm in the case of sparse matrices. The current configurations were chosen based on the chip model reconstructed with single photons via phase shifter calibrations (see \cite{Ilya} for details). The resulting transfer matrices represent typical two classes of transfer matrices that feature different sensitivity to the quality of experimental estimation of the visibility values. The matrices with smaller elements tend to render the corresponding interference visibility values hard to estimate because the corresponding events are detected rarely.  

We use matrix fidelity \(F\) as the distance metric between matrices:

\begin{equation}
    F(A, B) = \frac{Tr(A^\dagger B)Tr(B^\dagger A)}{Tr(A^\dagger A)Tr(A^\dagger A)},
\end{equation}
where \(Tr(A)\) is the trace of the matrix. Since fidelity is not preserved under a change of basis, and both unitary matrices are reconstructed up to phase factors in the input and output modes, it is necessary to bring both matrices to the same form for comparison. Additionally, in formula \ref{Visib} for visibility, matrices \(U\) and \(U^*\) will give the same experimental results (see appendix \ref{Cor_func_math}). Therefore, the matrix is reconstructed up to the sign of all phases. The matrices brought to the same form with real fourth row and second column in Fig. \ref{Exp_results}(a-b) have a fidelity \(F = 99.8\%\), and in Fig. 4 \(F = 98.8\%\).

\subsection{Boson Sampling}

After reconstructing the unitary matrix of the chip, we performed the two-photon boson sampling experiments. The single photons were injected into different pairs of input channels, and the coincidences were recorded between each pair of output channels. The single counts in each of the output channels were also stored. Then, simultaneously for all combinations of two photons at the input, the transmission coefficients \(t_i\), \(t'_j\), as well as the source parameters (indistinguishability $I$, emission probability \(p_{emit}\)) were optimized to reconstruct the obtained distributions at the outputs of the interferometer. Its matrix was considered to be known from the previous step. Using this method, we were able to reconstruct not only the coincidence distribution for all sets of input states but also the single-count distributions (see Fig. \ref{Exp_results} (c-d)). The average classical fidelity ($F_c = \sum_i\sqrt{p_iq_i}$) across all distributions was \(F_c = 99.6 \pm 0.3\%\). The obtained source parameters corresponded to those measured independently.
\section{Discussion}

While our method offers significant advantages, such as reduced measurement complexity and robustness to noise, it is important to acknowledge some potential limitations and disadvantages. The method relies on the quality of the single-photon source, particularly photon indistinguishability. While our approach is robust to some degree of imperfection, extremely low source efficiency or poor indistinguishability could degrade the accuracy of the reconstruction. Future improvements in single-photon source technology could mitigate this limitation. Although our method reduces the number of measurements compared to traditional approaches, it still requires a significant number of cross-correlation measurements, especially for larger interferometers. This could pose practical challenges in experimental setups with limited resources or time constraints. The global optimization algorithm used for reconstructing the transfer matrix requires a good initial guess to converge to the correct solution. In some cases, finding an appropriate initial guess might be non-trivial, potentially leading to suboptimal reconstructions. Further research into more robust optimization techniques could address this issue.

While our method shows promise for scalability with the number of modes, practical implementation in very large interferometers (e.g., with hundreds of modes) may still face challenges. The computational complexity of the optimization process and the need for precise control over a large number of parameters could become limiting factors. Although our method is robust to a certain level of noise, extremely high levels of experimental noise could still affect the accuracy of the reconstruction. Developing noise mitigation strategies or more sophisticated error correction techniques could further improve the method's reliability.

\section{Conclusion}
In this work, we have developed a novel method for reconstructing the transfer matrix of a linear-optical interferometer using cross-correlation functions of photon counts between pairs of output modes. Our approach is robust to experimental imperfections, such as losses and photon indistinguishability, and minimizes the requirements for input states, simplifying experimental implementation. We demonstrated the effectiveness of our technique through theoretical modeling and experimental validation in a 4-mode programmable integrated interferometer, achieving high fidelity in matrix reconstruction and successful application in boson sampling experiments. The method's scalability with the number of modes and its robustness to measurement errors make it a practical solution for characterizing linear-optical networks in quantum technologies.

\section{Acknowledgments}
The work was supported by Russian Science Foundation grant 22-12-00353-П (https://rscf.ru/en/project/22-12-00353/). The work was also supported by Rosatom in the framework of the Roadmap for Quantum computing (Contract No. 868-1.3-15/15-2021 dated October 5, 2021) in parts of Quantum Dot fabrication (Contract No. R2152 dated November 19, 2021) and demultiplexed single photon source development, photonic chip fabrication and calibration (Contract No.P2154 dated November 24, 2021). S.P.K. acknowledges support by Ministry of Science and Higher Education of the Russian Federation and South Ural State University (agreement №075-15-2022-1116). Yu.A. B is grateful to the Russian Foundation for the Advancement of Theoretical Physics and Mathematics (BASIS) (Projects №24-2-10-57-1).
\printbibliography
\appendix
\section{Correlation function formalism}
\label{Cor_func_math}
In general, the areas of the peaks on the measured cross-correlation histogram are determined by the correlation function of photon counts in the output modes of the interferometer:
\begin{equation}
    G_{demux}^{(2)}(t_1, t_2) = \langle a_4^\dagger(t_2)a_3^\dagger(t_1)a_4(t_2)a_3(t_1)\rangle
\end{equation}
To calculate its value, one can use the Heisenberg representation: express the annihilation operators in the output modes of the interferometer \(a_3, a_4\) in terms of the input operators \(a_1, a_2\) and calculate all expectation values over the input state.

We assume that the photon source emits with a period \(\tau_p\) (much larger than \(\gamma\)—the excitation lifetime in the QD) a state described by the density matrix \(\hat{\rho}(k)\) (the polarization degree of freedom is not considered due to the identical polarization of the considered photons, as well as the spatial distribution of the radiation due to the single-mode nature of the optical fibers and waveguides constituting the interferometer), where \(k\) is the number of the pump pulse exciting the source:
\begin{align}
\begin{split}
    &\hat{\rho}(k) = (1 -p_{emit})|vac\rangle\langle vac| + \\ 
    &(p_{emit} - p_{mp})\left(\iint_{-\infty}^{\infty}\rho(t_1 - k\tau_p, t_2 - k\tau_p)a^\dagger(t_1)|vac\rangle\langle vac|a(t_2)dt_1dt_2 + ...\right)
\end{split}    
\end{align}
where the multiphoton component is omitted, $p_{emit}$ -- single photon emission probability, $p_{mp}$ -- probability to emit multiple photons \(\rho(t_1, t_2) = \rho(t_2, t_1)^*\) due to the Hermiticity of the density matrix. That is, the source emits the same state at different times, tied to the arrival time of the pump pulse. Furthermore, the support of the function \(\rho(t_1, t_2)\) is limited to the region \([- \tau_p/2; \tau_p/2] \times [-\tau_p/2; \tau_p/2]\), i.e., \(\rho(t_1, t_2)\rho(t_1 - \tau_p, t_2 - \tau_p) = 0, \forall t_1, t_2\).

\begin{figure}[ht]
\centering
\includegraphics*[width=1.\linewidth]{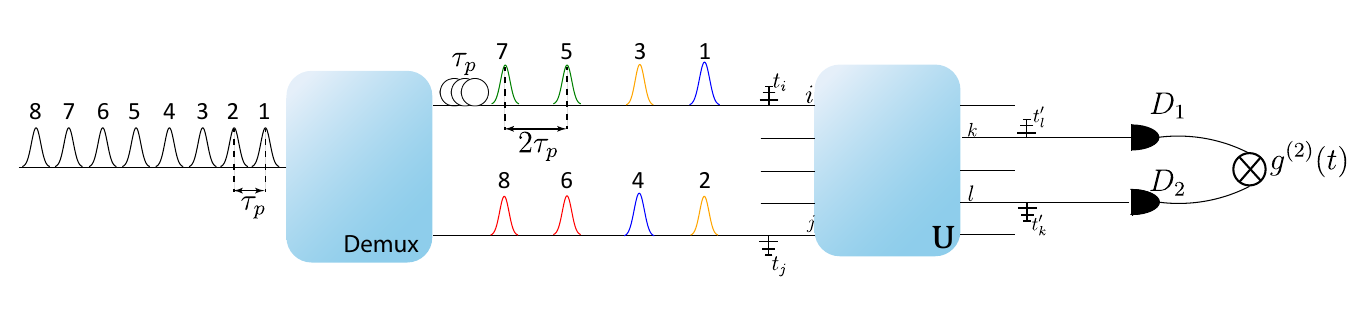}
\caption{Scheme of the experimental used for transfer matrix reconstruction. Demux - active demultiplexer routing single photons to the inputs of the interferometer. $D_1, D_2$ -- single photon detectors, $\tau_p$ - pump pulses period, $U$ -- transfer matrix of the interferometer.}
\label{Setup_demux}
\end{figure}

To implement transfer matrix reconstruction, we need to send a pair of photons at the input ports of the interferometer simultaneously. Fabrication of a set of identical QD sources is a notoriously hard task, so there is the task of using a single QD as a multiphoton source. We used an active two-channel demultiplexer based on a Pockels cell (see Fig. \ref{Setup_demux}). Further derivations will be conducted for this case, but it will be clear that they will be the same for the case of the photons generated simultaneously by different sources.

Using the active demultiplexer, this state is fed into the input modes \(i, j\) of the interferometer, and the cross-correlation of photon counts is calculated between the signals in the output modes \(k, l\). Since all interference effects are described by the matrix \(M'\) introduced in the main text, we will consider without loss of generality that the active demultiplexer sends states emitted by the quantum dot into two input modes 1, 2 of a non-unitary 2x2 transformer with transfer matrix \(M'\), between whose outputs the cross-correlation function is measured. If the histogram is taken over \(2N\) pump pulses, the total state at the input of the interferometer, taking into account the limited support of \(\rho(t_1, t_2)\), is described as follows:
\begin{equation}
\begin{split}
    \hat{\rho}_{total} =\left[\rho(1)_1\otimes\rho(1)_2\right] \otimes \left[\rho(3)_1\otimes\rho(3)_2\right]... = \\ 
    \otimes_{k=0}^{N-1}\left\{\left[\rho(2k+1)_1\otimes\rho(2k+1)_2\right]\right\},      
\end{split}
\end{equation}
where the index \(i\) in \(\hat{\rho}(2k)_i\) is the number of the input mode, and the number in parentheses numbers the temporal modes. Thus, at odd moments in time, due to the need to delay one of the single-photon pulses, no signal will be fed into the interferometer.
The annihilation operators in the output modes of the interferometer are obtained from the input operators by the transformation $M'$:

\begin{equation}
\begin{pmatrix}
    a_3(t_1) \\
    a_4(t_2) \\
\end{pmatrix} = 
\begin{pmatrix}
    M'_{11}a_1(t_1) + M'_{12}a_2(t_1)\\
    M'_{21}a_1(t_2) + M'_{22}a_2(t_2)\\
\end{pmatrix}
\end{equation}

Then the expression for the correlation function \(G^{(2)}_{demux}(t_1, t_2)\) taking into account the uncorrelatedness of states in different input spatial modes (e.g., $\langle a^\dagger_1(t_1)a_2(t_2)\rangle = \langle a^\dagger_1(t_1)\rangle \langle a_2(t_2)\rangle$) has the form:

\begin{align}
\label{correlation_gen}
    \begin{split}
        G_{demux}^{(2)}(t_1, t_2) = \left(|M'_{11}|^2|M'_{21}|^2 + |M'_{12}|^2|M'_{22}|^2\right)G^{(2)}(t_1,t_2)+\\
        \left(|M'_{11}|^2|M'_{22}|^2 + |M'_{12}|^2|M'_{21}|^2\right)I(t_1)I(t_2) + \\
        2Re\left\{M'^*_{11}M'^*_{22}M'_{12}M'_{21}\right\} |G^{(1)}(t_1,t_2)|^2 + \\ 2Re\left\{M'^*_{12}M'^*_{22}M'_{11}M'_{21}\right\}|C^{(2)}(t_1,t_2)|^2 +\\
        2Re\left\{\left(|M'_{11}|^2M'^*_{21}M'_{22} + |M'_{12}|^2M'^*_{22}M'_{21}\right)B_0(t_2, t_1)\right\} + \\ 
        2Re\left\{\left(|M'_{22}|^2M'^*_{12}M'_{11} + |M'_{21}|^2M'^*_{11}M'_{12}\right)B_0(t_1, t_2)\right\},
    \end{split}
\end{align}
where
\begin{align}
\begin{split}
    &G^{(2)}(t_1,t_2) = \langle a^\dagger(t_1)a^\dagger(t_2)a(t_1)a(t_2)\rangle - \text{multiphoton component,}\\
    &I(t_1) = \langle a^\dagger(t_1)a(t_1) \rangle - \text{Mean photon number,} \\
    &G^{(1)}(t_1, t_2) = \langle a^\dagger(t_2)a(t_1)\rangle - \text{First order correlation function,}\\
    &C^{(2)}(t_1, t_2) = \langle a(t_2)a(t_1)\rangle - \text{Second order coherence,}\\
    &B_0(t_1, t_2) = \langle a(t_2)\rangle \langle a^\dagger(t_2)a^\dagger(t_1)a(t_1) \rangle - \text{First order coherence}.
\end{split}
\end{align}
The quantity \(G^{(2)}(t_1, t_2)\) characterizes the source and is independently measured in the Hanbury-Brown-Twiss scheme; \(G^{(1)}(t_1, t_2)\) is directly related to the average overlap of wave packets emitted by the QD at different times, i.e., to the indistinguishability \(I\):
\begin{equation}
    I = \frac{1}{\mu^2}\iint |G^{(1)}(t_1, t_2)|^2 dt_1dt_2,
\end{equation}

where \(\mu = \int I(t)dt\) is the average number of photons entering the interferometer from one channel. \(C^{(2)}(t_1, t_2)\) depends on the off-diagonal elements \(\rho_{n,n-2}\) of the density matrix in the Fock basis, responsible for coherence between Fock states differing by 2 in photon number. \(B(t_1, t_2)\) is determined by the elements \(\rho_{n,n-1}\) and is a manifestation of single-photon interference effects on the cross-correlation histogram.

Note that replacing matrix \(M'\) with \(M'^*\) does not change the form of formula \ref{correlation_gen}. Therefore, the reconstructed interferometer matrix is determined up to the sign of all phases.

In the case of considering the side peaks on the cross-correlation histogram, i.e., under the condition \(|t_2 - t_1| > \tau_p\), most terms in expression \ref{correlation_gen} will take a simpler form due to the uncorrelatedness of states emitted by the source at different moments of time.

\begin{align}
    \begin{split}
        G_{demux, uncor}^{(2)}(t_1, t_2) = \left(|M'_{11}|^2 + |M'_{12}|^2\right) \left(|M'_{21}|^2 + |M'_{22}|^2\right) I(t_1)I(t_2) + \\
        2Re\left\{M'^*_{11}M'^*_{22}M'_{12}M'_{21} + M'^*_{12}M'^*_{22}M'_{11}M'_{21} \right\} |C^{(1)}(t_1)|^2|C^{(1)}(t_2)|^2 + \\
        2Re\left\{M'^*_{21}M'_{22}\right\}\left(|M'_{11}|^2 + |M'_{12}|^2\right)I(t_2)|C^{(1)}(t_1)|^2 + \\
        2Re\left\{M'^*_{12}M'_{11} \right\}\left(|M'_{22}|^2 + |M'_{21}|^2\right) I(t_1)|C^{(1)}(t_2)|^2,
    \end{split}
\end{align}
где $C^{(1)}(t) = \langle a(t) \rangle$ is also photonic first order coherence related to matrix elements $\rho_{n, n-1}$.

To obtain the areas of the peaks on the cross-correlation histograms, it is necessary to integrate the obtained expressions for the correlation functions over time \( t_1, t_2 \). Additionally, for convenience, we normalize the integrated expressions by \( \mu^2 \)—the value of the cross-correlation function in the absence of interference and correlation between the two channels. For the central peak, we obtain:

\begin{align}
    \begin{split}
        &g_{demux}^{(2)} = \left(|M'_{11}|^2|M'_{22}|^2 + |M'_{12}|^2|M'_{21}|^2\right) +\\ &\left(|M'_{11}|^2|M'_{21}|^2 + |M'_{12}|^2|M'_{22}|^2\right)g^{(2)}(0)  +\\
        &2Re\left\{M'^*_{11}M'^*_{22}M'_{12}M'_{21}\right\}I +2Re\left\{M'^*_{12}M'^*_{22}M'_{11}M'_{21}\right\}c^{(2)} +\\
        &2Re\left\{\left(M'_{11}M'_{22} + M'_{12}M'_{21}\right)\left(M'^*_{21}M'^*_{11} + M'^*_{22}M'^*_{12}\right) 
        b_0\right\},
    \end{split}
\end{align}
where
\begin{align}
    \begin{split}
        g^{(2)}(0) = \frac{1}{\mu^2}\iint G^{(2)}(t_1, t_2)dt_1dt_2,\\
        c^{(2)} = \frac{1}{\mu^2}\iint |C^{(2)}(t_1, t_2)|^2dt_1dt_2,\\
        b_0 = \frac{1}{\mu^2}\iint B_0^{(2)}(t_1, t_2)dt_1dt_2.
    \end{split}
\end{align}
For the side peaks, we similarly obtain:
\begin{align}
    \begin{split}
        &g_{demux, uncor}^{(2)} = \left(|M'_{11}|^2 + |M'_{12}|^2\right) \left(|M'_{21}|^2 + |M'_{22}|^2\right)+ 4Re\left\{M'^*_{22}M'_{21}\right\}Re\left\{M'^*_{11}M'_{12}\right\}{c^{(1)}}^2+\\
        &\left[2Re\left\{M'^*_{21}M'_{22}\right\}\left(|M'_{11}|^2 + |M'_{12}|^2\right) + 2Re\left\{M'^*_{12}M'_{11} \right\}\left(|M'_{22}|^2 + |M'_{21}|^2\right)\right]c^{(1)} = \\
        &\left[\left(|M'_{11}|^2 + |M'_{12}|^2\right) + 2Re\left\{M'^*_{11}M'_{12}\right\}c^{(1)}\right] \left[\left(|M'_{21}|^2 + |M'_{22}|^2\right) + 2Re\left\{M'^*_{21}M'_{22}\right\}c^{(1)}\right],\\
        &\text{where }c^{(1)} = \frac{1}{\mu}\int |C^{(1)}(t)|^2dt.
    \end{split}
    \label{uncor_g2}
\end{align}

If we neglect all effects related to photon coherence and the multiphoton component in the state \(\hat{\rho}\) (\(b_0 = c^{(1)} = c^{(2)} = g^{(2)}(0) = 0\)), i.e., transition to the picture of independent single-photon pulses considered in the main text, it is easy to see that the ratio of the central peak area to the side peak transitions into formula \ref{Visib}. The use of this approximation in reconstructing the transfer matrix of the interferometer is justified because the photon source based on a quantum dot in a microresonator, used in our setup, was pumped in the resonant regime near the \(\pi\)-pulse, i.e., all off-diagonal components of the density matrix are close to zero at the moment of emission and could only decrease due to losses in the photon collection, demultiplexing, and coupling into the interferometer. Additionally, the source itself has a low multiphoton component: the measured value of \(g^{(2)}(0) \approx 0.03\). All of this allowed us to describe the interference visibility introduced in the main text with good accuracy and, ultimately, to reconstruct the transfer matrix. Furthermore, simulation results show that the reconstruction procedure is robust to errors in visibility measurements up to \(20\%\) (see \ref{section: Simulation}) of the absolute value, which allows for the use of a more intuitive and straightforward representation of photon wave packets as independent pulses even in the presence of relatively small photon number coherence.

\section{Measurement of photon indistiguishability in active demultiplexer setup}
The procedure for measuring the indistinguishability \(I\), introduced in Section \ref{Cor_func_math}, is based on measuring the visibility of Hong-Ou-Mandel interference and is mathematically described in the same way as measuring the cross-correlation function in the tomography of a multimode interferometer and is essentially a special case of this measurement.

\begin{figure}[H]
\centering
\includegraphics*[width=.75\linewidth]{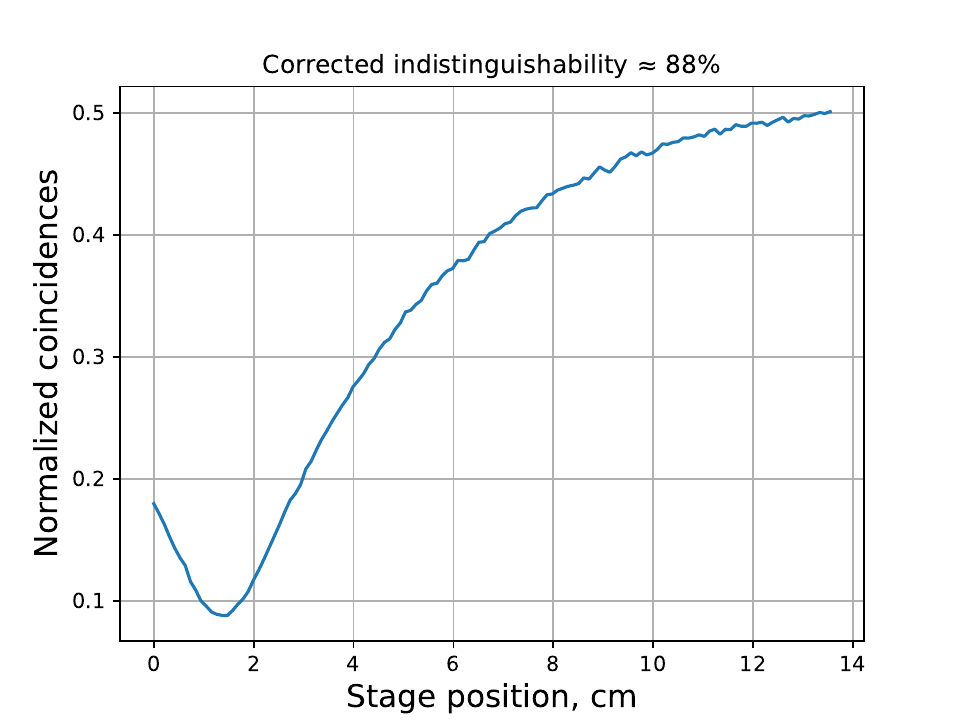}
\caption{Mandel dip for photons from the QD emitted with an interval of $12.1$ ns.}
\label{HOM_dip}
\end{figure}

The matrix \(M'\), introduced in the main part and actively used in Section \ref{Cor_func_math}, will now coincide with matrix \(M\) and take a specific form. We will use the following parameterization of the beam splitter:

\begin{equation}
    U_{BS} = \begin{pmatrix}
        \sqrt{R} & \sqrt{T}\\
        \sqrt{T} & -\sqrt{R}
    \end{pmatrix},
    R + T = 1.
\end{equation}
Then matrix \(M'\) takes the form:
\begin{equation}
    M' = \begin{pmatrix}
        \sqrt{\eta_1\eta'_1}\sqrt{R} & \sqrt{\eta_2\eta'_1}\sqrt{T}\\
        \sqrt{\eta_1\eta'_2}\sqrt{T} & -\sqrt{\eta_2\eta'_2}\sqrt{R}
    \end{pmatrix}
\end{equation}
First, consider the special case where photon coherence effects can be neglected (\(c^{(1)} = b_0 = c^{(2)} = 0\)), which corresponds to resonant pumping near the \(\pi\)-pulse. We assume that both input transmissions \(\eta_1, \eta_2\) are non-zero, otherwise the indistinguishability measurement scheme turns into a multiphoton component measurement scheme \(g^{(2)}(0)\). In this case, the expressions for the areas of the central and side peaks take the form:
\begin{align}
    \begin{split}
        &\frac{g^{(2)}_{HOM}}{\eta_1\eta_2\eta'_1\eta'_2} = \left(R^2 + T^2\right) +
        (r_\eta + \frac{1}{r_\eta})RTg^{(2)}(0) - 2RTI\\
        &\frac{g^{(2)}_{HOM, uncor}}{\eta_1\eta_2\eta'_1\eta'_2} = \left(R + r_\eta T\right)\left(\frac{1}{r_\eta} T + R\right) = 1 + \left(\sqrt{r_\eta} - \sqrt{\frac{1}{r_\eta}}\right)^2RT
    \end{split}
\end{align}
here the ratio of input efficiencies \(r_\eta = \eta_2/\eta_1\) is introduced, which can be easily determined experimentally, as well as the splitting coefficients \(R, T\) \cite{laing2012super}. Then, introducing \(V = g^{(2)}_{HOM}/g^{(2)}_{HOM,uncor}\), we obtain the expression for indistinguishability:
\begin{equation}
    I = \frac{R^2 + T^2}{2RT} + \frac{1}{2}(r_\eta + \frac{1}{r_\eta})g^{(2)}(0) - \left(\frac{1}{2RT} + \frac{1}{2}\left(\sqrt{r_\eta} - \sqrt{\frac{1}{r_\eta}}\right)^2\right)V, 0 < r_\eta < \infty. 
    \label{Indisting_formula}
    \end{equation}
In the indistinguishability measurement experiment, the same two-channel demultiplexer was used as in the interferometer matrix measurements. One of the input channels of the beam splitter was placed on a motorized platform, which allowed introducing a delay between the arrival times of photons at the beam splitter, thus regulating their indistinguishability. As a result, it was possible to observe the Mandel dip \cite{hong1987measurement} for photons emitted by the QD at successive times (see Fig. \ref{HOM_dip}). 
The obtained value of indistinguishability, corrected using formula \ref{Indisting_formula}, was 88\%, and the photon coherence length was about 3 cm. To our knowledge, it is the first time when Mandel dip is measured for photons with such a long coherence length.

\end{document}